\documentclass[12pt]{article}
\usepackage[english]{babel}
\usepackage[latin1]{inputenc}

\usepackage[intlimits]{amsmath}
\usepackage{amssymb}

\usepackage{url}
\usepackage{graphicx}
\usepackage{slashed}
\usepackage{color}
\usepackage{axodraw}
\usepackage[text={15.9cm,21.7cm}]{geometry}

\begin{document}


\title{\vspace{-2cm} 
{\normalsize
\flushright TUM-HEP 851/12\\
\vspace{-0.5cm} 
\flushright FLAVOUR(267104)-ERC-24\\}
\vspace{0.8cm} 
\bf A radiative model with a naturally mild neutrino mass hierarchy\\[8mm]}

\author{Dieter Hehn, Alejandro Ibarra\\[2mm]
{\normalsize\it Physik-Department T30d, Technische Universit\"at M\"unchen,}\\[-0.05cm]
{\it\normalsize James-Franck-Stra\ss{}e, 85748 Garching, Germany}
}

\maketitle

\begin{abstract}
Many neutrino mass models postulate the existence of at least two extra fermions in order to account for the measured solar and atmospheric mass splittings. In these models, however, the predicted hierarchy between the two mass splittings is generically much larger than the observed one, unless extra flavor symmetries are introduced. We present in this letter a radiative neutrino mass model consisting of the Standard Model extended by one heavy fermionic singlet and two scalars which predicts, under very general conditions, a neutrino mass hierarchy in qualitative agreement with the experimental value. 
\end{abstract}

\maketitle

\section{Introduction}

Unraveling the origin of neutrino masses and understanding the striking differences between neutrino parameters and quark parameters is one of the major challenges in theoretical Particle Physics. Over the last thirty years various schemes have been proposed which can plausibly explain the smallness of neutrino masses, either at tree level or via quantum effects. However, accommodating simultaneously in these schemes the observed large mixing angles and the mild hierarchy between the atmospheric and solar mass splittings has proven to be a much more difficult task.

Mechanisms of neutrino mass generation at tree level can be classified in three types: type I, II and III through the introduction of heavy fermion singlets \cite{typeI}, heavy scalar triplets \cite{typeII} or heavy fermion triplets \cite{typeIII}, respectively. In tree level models introducing extra fermions, at least two generations of them must be postulated in order to account for the observed atmospheric and solar neutrino mass splittings. Under the plausible assumption that the masses of the extra fermions present an intergenerational mass hierarchy, as observed in the quark and charged lepton sectors, a very large hierarchy is generically expected between the atmospheric and solar mass splittings~\cite{Casas:2006hf}. Possible ways out consist in assuming very mild hierarchies in the right-handed neutrino masses {\it and} in the neutrino Yukawa eigenvalues or, alternatively, assuming a very large hierarchy in the right-handed neutrino masses, roughly quadratic with respect to the hierarchy in the neutrino Yukawa eigenvalues, {\it and} small mixing angles in the right-handed sector, so that in the effective theory both hierarchies almost exactly compensate each other. While these conditions seem bizarre from the phenomenological point of view, they can be fulfilled in concrete flavor models (see {\it e.g.} \cite{Buchmuller:2011tm}). This drawback does not exist, though, in the type II see-saw model, where the effective neutrino mass matrix is simply proportional to the Yukawa coupling of the scalar triplet to the lepton doublets. Hence, postulating a mild hierarchy in the Yukawa eigenvalues directly translates into a mild hierarchy in the neutrino masses. 

Radiative models of neutrino masses introducing extra fermions generically face the same challenges as tree level models when trying to explain the mild neutrino mass hierarchy. An interesting exception arises when the Standard Model is extended with a second Higgs doublet and with (at least) one heavy fermion singlet. In this model, the atmospheric mass scale arises at tree level, while the solar mass scale at the one loop level \cite{Ibarra:2011gn,Grimus:1999wm}, being the neutrino mass hierarchy proportional to a loop factor times a large logarithm of the ratio between the right-handed neutrino mass and the Higgs mass, giving naturally $m_2/m_3~\sim {\cal O}(0.1)$, in qualitative agreement with experiments. Furthermore, this scenario has the attractive feature that the resulting mass hierarchy is fairly insensitive to the concrete values of the new mass scales introduced in the theory. In particular, the same conclusion holds when all the extra particles are very heavy, thus suppressing their potentially dangerous effects in flavor and CP violating processes as well as in the electroweak precision observables \cite{Ibarra:2011gn}.

In this paper we will explore the possibility of explaining the observed mild neutrino mass hierarchy in models with just radiative neutrino mass generation. Our starting point will be the model first considered by Ma~\cite{Ma:2006km}, consisting of the Standard Model extended by one scalar doublet, $\eta$, with the same gauge quantum numbers as the Higgs doublet and several fermionic gauge singlets. It is also postulated that all the new particles are odd under a discrete $Z_2$ symmetry, while the Standard Model particles are even (other radiative models were proposed, {\it e.g.}, in \cite{Witten:1979nr,Zee:1980ai,Wolfenstein:1980sy,Zee:1985id,Babu:1988ki,Babu:1989pz,Aoki:2008av}). This model not only generates small neutrino masses, but it also provides a candidate for dark matter as well as a rich collider phenomenology. However, following the general arguments outlined above, this model tends to generate a neutrino mass hierarchy much larger than the observed one. We will show in this letter that by adding a second scalar particle with identical gauge and discrete charges as $\eta$ a mild neutrino mass hierarchy naturally arises, while preserving the remaining features of the model.

\section{Model}

We consider a model where the particle content of the Standard Model is extended by one extra Majorana fermion $\chi$, singlet under the Standard Model gauge group and two complex scalars $\eta_{1,2}$, doublets under $SU(2)_L$ and with hypercharge $1/2$. Furthermore, we postulate the existence of a discrete $Z_2$ symmetry, under which the Standard Model particles are even while the new particles $\chi$, $\eta_{1,2}$ are odd. With this particle content, the most general Lagrangian reads:
\begin{equation}
{\cal L}={\cal L}_{\rm SM}+{\cal L}_{\chi}+{\cal L}_\eta+
{\cal L}^{\rm fermion}_{\rm int}+{\cal L}^{\rm scalar}_{\rm int}\;.
\end{equation}
Here, ${\cal L}_{\rm SM}$ is the Standard Model Lagrangian which includes a potential for the Higgs doublet $\Phi$, $V=m_\Phi^2 \Phi^\dagger \Phi +\frac{1}{2}\lambda_\Phi (\Phi^\dagger \Phi)^2$. On the other hand ${\cal L}_\chi$ and ${\cal L}_\eta$ are the parts of the Lagrangian involving just the Majorana fermion $\chi$ and the scalar particles $\eta_{1,2}$, respectively, and which are given by
\begin{eqnarray}
{\cal L}_\chi&=&\frac{1}{2} \overline{\chi^c} i\slashed{\partial} \chi -\frac{1}{2} M \overline{\chi^c}\chi +{\rm h.c.}\;, \\
{\cal L}_\eta&=&(D_\mu \eta^a)^\dagger  (D^\mu \eta_a)-(m_\eta^2)_{ab} \eta^{a\dagger}\eta^b- V(\eta_1,\eta_2)\;,
\end{eqnarray}
where $D_\mu$ denotes the covariant derivative and $V(\eta_1,\eta_2)$ is the general potential of a two Higgs doublet model.  Note that $\eta_1$ and $\eta_2$ have identical quantum numbers, hence internal rotations between them leave the action invariant. Therefore, in what follows we will define these fields such that the mass matrix $(m^2_\eta)_{ab}$ is diagonal with entries $m_{\eta_1}^2$ and $m_{\eta_2}^2$. Lastly, ${\cal L}^{\rm fermion}_{\rm int}$ and ${\cal L}^{\rm scalar}_{\rm int}$ denote the fermionic and scalar interaction terms of the new particles to the left-handed leptons, $L_i=(\nu_i, \ell^-_i)$, and to the Higgs doublet, $\Phi$:
\begin{eqnarray}
\begin{split}
{\cal L}^{\rm fermion}_{\rm int}
 =&-Y^{(a)}_i \bar \chi (L_i i\sigma_2 \eta_a)+{\rm h.c.} 
 =-Y^{(a)}_i \bar \chi (\nu_i \eta_a^0-\ell^-_i \eta_a^+)+{\rm h.c.}\;,  \\
{\cal L}^{\rm scalar}_{\rm int}=&
-\frac{1}{2}\lambda_3^{(ab)}(\Phi^\dagger \Phi)(\eta_a^\dagger \eta_b)
-\frac{1}{2}\lambda_4^{(ab)}(\Phi^\dagger \eta_a)(\eta_b^\dagger \Phi) 
-\frac{1}{2}\lambda_5^{(ab)}(\Phi^\dagger \eta_a)(\Phi^\dagger \eta_b)+{\rm h.c.}
\end{split}
\end{eqnarray}

Due to the discrete $Z_2$ symmetry the fields $\eta_1$ and $\eta_2$ do not acquire a vacuum expectation value, while the field $\Phi$ does, along its neutral direction $\langle \Phi^0 \rangle\equiv v/\sqrt{2}$. After the electroweak symmetry breaking there are nine physical degrees of freedom in the scalar sector, the Standard Model Higgs, $h$, four exotic charged scalars and four exotic neutral scalars. Due to the interaction terms $\lambda_{3,4,5}^{(ab)}$ the neutral scalar mass matrix has a complicated structure leading to four different mass eigenstates. We will assume for simplicity that $\lambda_{3,4,5}^{(ab)} v^2\ll (m^2_\eta)_{11},(m^2_\eta)_{22}$, hence there are two electrically neutral degrees of freedom with mass $m_{\eta_1}$ and another two with mass $m_{\eta_2}$. Without loss of generality, we will label the mass eigenstates such that  $m_{\eta_1}<m_{\eta_2}$.

\begin{figure}
\begin{center}
\begin{picture}(0,60)(80,0)
\ArrowLine(0,0)(40,0)
\ArrowLine(40,0)(80,0)
\ArrowLine(120,0)(80,0)
\ArrowLine(160,0)(120,0)
\DashArrowArcn(80,0)(40,90,0){4}
\DashArrowArc(80,0)(40,90,180){4}
\DashArrowLine(110,70)(80,40){4}
\DashArrowLine(50,70)(80,40){4}
\Text(20,-10)[c]{$\nu_i$}
\Text(140,-10)[c]{$\nu_j$}
\Text(80,-10)[c]{$\chi$}
\Text(80,0)[c]{$\times$}
\Text(38,25)[c]{$\eta^0_a$}
\Text(122,25)[c]{$\eta^0_b$}
\Text(50,60)[c]{$\Phi^0$}
\Text(110,60)[c]{$\Phi^0$}
\end{picture}
\end{center}
\caption{Diagrams contributing to the effective neutrino mass matrix ${\cal M}_\nu$, with $a,b=1,2$.}
\label{diagram}
\end{figure}
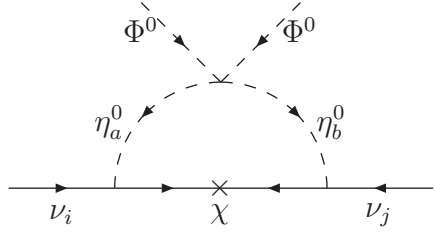

Since there is no interaction term between the Majorana fermion $\chi$ and the Standard Model Higgs doublet $\Phi$, no tree level mass for the neutrinos is generated. However, at the one loop level, a neutrino mass matrix is generated through the diagram in Fig.~\ref{diagram}, the result being
\begin{equation}
({\cal M}_\nu)_{ij}=\frac{Y_i^{(a)} Y_j^{(b)} \lambda_5^{(ab)} v^2}{8\pi^2}\frac{M}{m_{\eta_b}^2-M^2} \left\{\frac{m_{\eta_b}^2}{m_{\eta_a}^2-m_{\eta_b}^2}\log \frac{m_{\eta_a}^2}{m_{\eta_b}^2}-\frac{M^2}{m_{\eta_a}^2-M^2}\log\frac{m_{\eta_a}^2}{M^2}\right\}\;,
\label{eq:master}
\end{equation}
where a sum over $a,b=1,2$ is implicitly assumed.

We can analyze this mass matrix in different interesting limits. First, in the case that one of the extra scalars, say $\eta_2$, decouples, either because $m_{\eta_2}\gg m_{\eta_1}, M$ or because $Y_j^{(2)}\ll Y_j^{(1)}$, eq.(\ref{eq:master}) reduces to the well known form \cite{Ma:2006km}
\begin{equation}
({\cal M}_\nu)_{ij}\simeq\frac{Y_i^{(1)} Y_j^{(1)} \lambda_5^{(11)} v^2}{8\pi^2}\frac{M}{m_{\eta_1}^2-M^2}  \left\{1-\frac{M^2}{m_{\eta_1}^2-M^2}\log\frac{m_{\eta_1}^2}{M^2}\right\}\;.
\end{equation}
On the other hand, in the limit $M \gg m_{\eta_1},m_{\eta_2}$ the mass matrix can be approximated by:
\begin{equation}
({\cal M}_\nu)_{ij}\simeq-\frac{Y_i^{(a)} Y_j^{(b)} \lambda_5^{(ab)} v^2}{8\pi^2}\frac{1}{M}\left\{\frac{m_{\eta_b}^2}{m_{\eta_a}^2-m_{\eta_b}^2}\log \frac{m_{\eta_a}^2}{m_{\eta_b}^2}+\log\frac{m_{\eta_a}^2}{M^2}\right\}\;,
\end{equation}
while in the limit $m_{\eta_1},m_{\eta_2} \gg M$ by
\begin{equation}
({\cal M}_\nu)_{ij}\simeq\frac{Y_i^{(a)} Y_j^{(b)} \lambda_5^{(ab)} v^2}{8\pi^2}
\frac{M}{m_{\eta_a}^2-m_{\eta_b}^2}\log \frac{m_{\eta_a}^2}{m_{\eta_b}^2}\;.
\end{equation}
In both cases, the smallness of neutrino masses can be explained by making the heaviest scale much larger than the electroweak symmetry breaking scale, or by making the couplings $Y^{(a)}_i$, $\lambda^{(ab)}$ small. This scenario is further constrained by the electroweak precision data, which require the contribution of the extra scalars to the oblique parameters to be small~\cite{Peskin:1990zt}. Concretely, the contribution of the extra scalars to the $T$ parameter can be suppressed when the custodial symmetry breaking parameters $\lambda^{(ab)}_4$ and $\lambda^{(ab)}_5$ are small, or when the extra scalars decouple, $m_{\eta_1,\eta_2}\gg v$. Lastly, if the $Z_2$ symmetry is exact, the lightest particle of the $Z_2$ odd sector is absolutely stable and will survive until today as a thermal relic. Hence, additional constraints on the parameters of the model follow from requiring a particle density equal to the observed dark matter density or, more conservatively, from avoiding dark matter overproduction. These constraints can be avoided, though, if the $Z_2$ symmetry is mildly broken. In this paper we will not consider the interesting possibility that the lightest between $\eta_1$ and $\chi$ could constitute the dark matter of our Universe, but we will just focus on the implications of the model for neutrino physics.

To explore the phenomenology of this model, it is convenient to cast eq.(\ref{eq:master}) in the following form:
\begin{equation}
({\cal M}_\nu)_{ij}=\tilde Y_i^{(c)} \Lambda_c \tilde Y_j^{(c)} \;.
\label{eq:mass2}
\end{equation}
Here $\tilde Y_i^{(c)}=W_{ac} Y^{(a)}$, where $W$ is the unitary matrix that diagonalizes the matrix
\begin{equation}
\Lambda^{(ab)}=\frac{\lambda_5^{(ab)} v^2}{8\pi^2}\frac{M}{m_{\eta_b}^2-M^2} 
\left\{\frac{m_{\eta_b}^2}{m_{\eta_a}^2-m_{\eta_b}^2}\log \frac{m_{\eta_a}^2}{m_{\eta_b}^2}-\frac{M^2}{m_{\eta_a}^2-M^2}\log\frac{m_{\eta_a}^2}{M^2}\right\}\;,
\label{eq:Lambda-master}
\end{equation}
through $\Lambda^{(ab)}=W_{ac} \Lambda_c W_{bc}$, and $\Lambda_c$ are the elements of the diagonalized matrix.

The neutrino mass eigenvalues can be straightforwardly calculated from eq.(\ref{eq:mass2}), the result being:
\begin{equation}
m^2_{3,2}=\frac{1}{2}\Big(t\pm \sqrt{t^2-4 d^2}\Big)\;,
\end{equation}
with
\begin{eqnarray}
t&=&\Lambda_1^2|\tilde Y^{(1)}|^4+\Lambda_2^2|\tilde Y^{(2)}|^4+2 \Lambda_1\Lambda_2{\rm Re}\Big[(\tilde Y^{(1)\dagger}\tilde Y^{(2)})^2\Big]\;, \\
d&=&\Lambda_1\Lambda_2 \Big(|\tilde Y^{(1)}|^2| \tilde Y^{(2)}|^2- |\tilde Y^{(1)\dagger} \tilde Y^{(2)}|^2\Big)\;.
\end{eqnarray}
In these expressions, $|\tilde Y^{(a)}|^2=\sum |\tilde Y_i^{(a)}|^2$ are the norms of the vector columns corresponding to the Yukawa coupling to the scalar $\eta_a$ and $\tilde Y^{(a)\dagger} Y^{(b)}= \sum_i \tilde Y_i^{(a)*}\tilde Y_i^{(b)}$. 

We expect, under very general conditions, the neutrino mass hierarchy $m_2/m_3$ to be mild. Given that the two scalar fields $\eta_1$ and $\eta_2$ have the same quantum numbers, it is reasonable to assume that the quartic couplings $\lambda_5^{(ab)}$ are all of the same order and that $m_{\eta_1}\sim m_{\eta_2}$, from where it follows that $\Lambda_1\sim \Lambda_2$. Besides, the vectors $\tilde Y^{(a)}$ are linear combinations of the Yukawa vectors $Y^{(a)}$, which in turn are plausibly having a similar norm $|Y^{(1)}|\sim |Y^{(2)}|$ and are misaligned. Therefore, we also expect $|\tilde Y^{(1)}|\sim |\tilde Y^{(2)}|$ and that $\tilde Y^{(1)}$ and $\tilde Y^{(2)}$ are  misaligned. From these two general considerations, it follows that $t\sim d$ and hence $m_2/m_3\sim 0.1-1$.

\begin{figure}[t]
\begin{center}
\includegraphics[width=7cm]{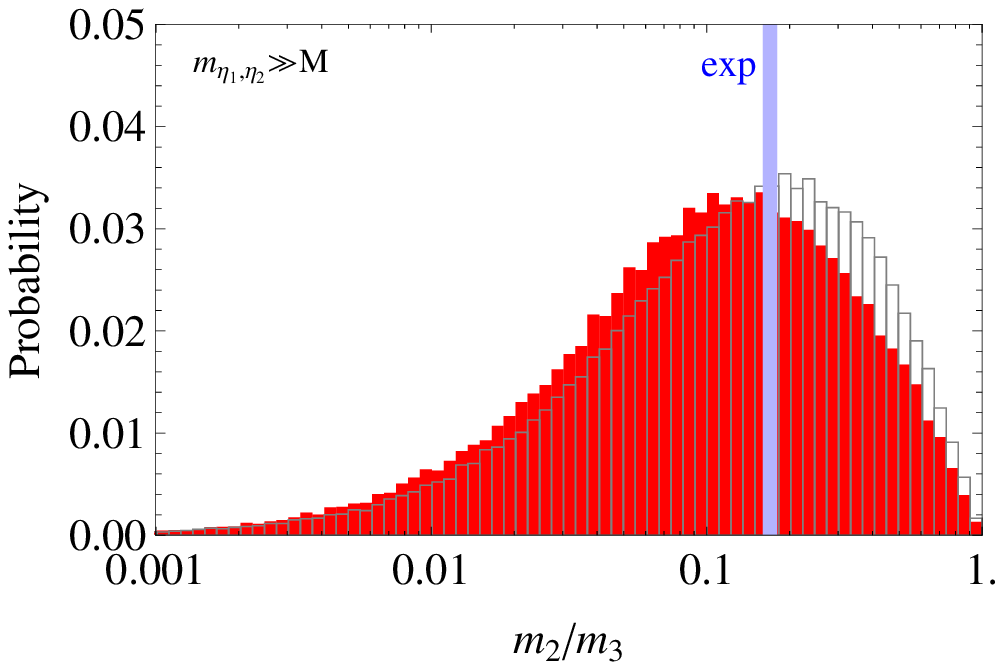}
\includegraphics[width=7cm]{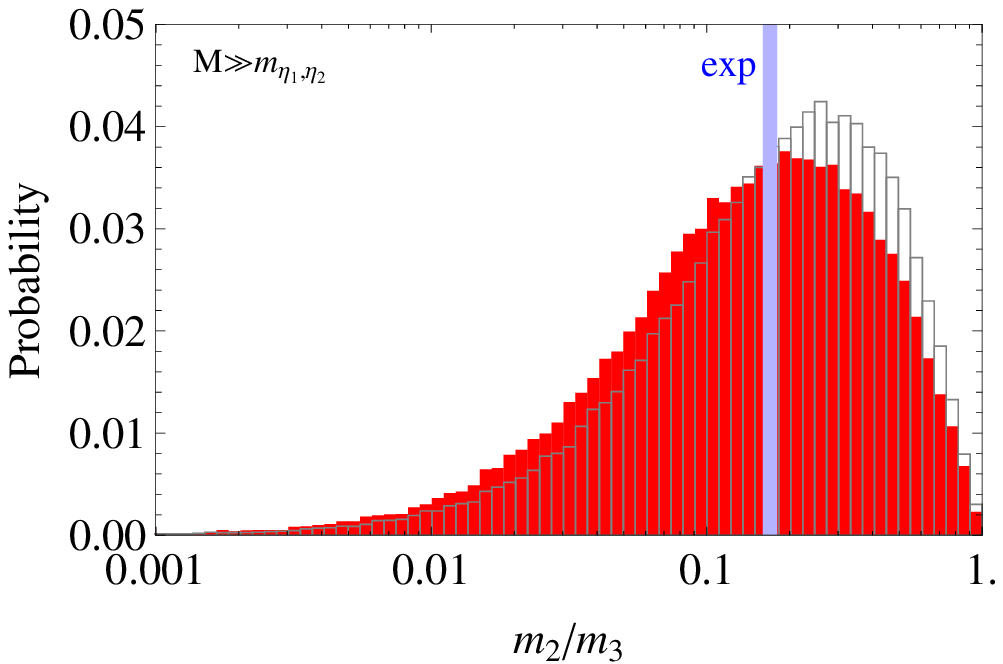}
\caption{Probability distributions of the mass ratio $m_2/m_3$ for the case $m_{\eta_1,\eta_2}\gg M$ (left plot) and $M\gg m_{\eta_1,\eta_2}$ (right plot) from a random scan of the parameters. Details are given in the text.
}\label{fig:1}
\end{center}
\end{figure}

This conclusion is confirmed by our numerical analysis. We show in Fig.~\ref{fig:1} the probability distributions of $m_2/m_3$ in a logarithmic binning from performing a random scan of the parameters entering the mass matrix for the two limits $m_{\eta_1,\eta_2}\gg M$ and $M\gg m_{\eta_1,\eta_2}$. We have taken flat distributions of the moduli and phases of the complex parameters $Y^{(a)}_i$ and $\lambda_5^{(ab)}$ in the range 0 to 1 and 0 to $2\pi$, respectively, and of the ratio between the two scalar masses $m_{\eta_2}/m_{\eta_1}$ in the range 1 to 3. Finally, in order to reproduce the correct value of the atmospheric mass splitting one can introduce an overall rescaling of the Yukawa couplings and/or the quartic couplings and/or the mass scales involved. The resulting mass ratio probability distributions are shown in the histograms as continuous lines, however, most of the points do not correctly reproduce the correct mixing angles. We show as red histograms the mass ratio probability distributions after imposing that the mixing angles lie within the $2\sigma$ ranges derived in the latest global fit to the neutrino oscillation parameters: $\sin^2\theta_{12}=0.29-0.35$, $\sin^2\theta_{23}=0.38-0.66$, $\sin^2\theta_{13}=0.019-0.030$ \cite{Tortola:2012te} (the global fit presented in \cite{Fogli:2012ua} finds similar $2\sigma$ ranges for $\sin^2\theta_{12}$ and $\sin^2\theta_{13}$ but a significanly narrower range for $\sin^2\theta_{23}=0.348-0.448$, which, nevertheless, does not affect our conclusions). Lastly, we also show for reference the experimental value of the ratio $m_2/m_3$, determined from the $2\sigma$ ranges $\Delta m^2_{\rm atm}=(2.38-2.68)\times 10^{-3}\,{\rm eV}^2$,  $\Delta m^2_{\rm sol}=(7.27-8.01)\times 10^{-5}\,{\rm eV}^2$ \cite{Tortola:2012te}.  As apparent from the plot, in this scenario large neutrino mass hierarchies are disfavored, while values around $m_2/m_3\sim 0.1$ are favored. 

It is also important to note that this mechanism to generate a mild neutrino mass hierarchy is uncorrelated to the pattern of neutrino mixing angles, as follows from the similarity of the probability distributions of the mass ratio when including or disregarding the requirement of reproducing the correct mixing angles. Therefore, in this framework constructing a successful model of neutrino masses only requires a compelling explanation for the observed pattern of neutrino mixing angles; reproducing the small neutrino masses is a natural consequence of introducing heavy scales (or small couplings) in the theory, and reproducing the mild neutrino mass hierarchy is a natural consequence of introducing the two scalar fields $\eta_1$ and $\eta_2$.

\section{Conclusions}

We have presented a neutrino mass model consisting of the Standard Model extended with a fermionic gauge singlet and two scalar particles with identical gauge quantum as the Standard Model Higgs doublet. All the new particles are assumed to be odd under a discrete $Z_2$ symmetry while the Standard Model particles are all even. Neutrino masses appear at the one loop level and are naturally small if the masses of the new particles are large and/or the new couplings are small. Furthermore, a mild neutrino mass hierarchy naturally arises since both neutrino masses are suppressed by the same scale. Hence, the mass ratio only depends on a complicated combination of the couplings and plausibly lies in the range $m_2/m_3=0.1-1$. No flavor symmetry is then necessary to explain the mass hierarchy between the different neutrino generations, although it might be necessary to explain the observed pattern of mixing angles. As in Ma's model, the model contains a candidate for dark matter and presents a rich phenomenology, which will be explored somewhere else.

\section*{Acknowledgements}
This work was partially supported by the DFG cluster of excellence ``Origin and Structure of the Universe.''


\end{document}